\documentclass[journal=ancac3,manuscript=letter,layout=twocolumn]{achemso}

\usepackage[version=3]{mhchem} 
\usepackage{graphicx}
\usepackage{dcolumn}
\usepackage{bm}
\usepackage[mathlines]{lineno}
\usepackage{amsmath}

\author{Jie Guan}
\altaffiliation{These authors contributed equally to this work.}
\affiliation{Physics and Astronomy Department,
             Michigan State University,
             East Lansing, Michigan 48824, USA}
\author{Zhen Zhu}
\altaffiliation{These authors contributed equally to this work.}
\affiliation{Physics and Astronomy Department,
             Michigan State University,
             East Lansing, Michigan 48824, USA}
\author{David Tom\'{a}nek}
\affiliation{Physics and Astronomy Department,
             Michigan State University,
             East Lansing, Michigan 48824, USA}
\email{tomanek@pa.msu.edu}

\title{\hfill {\small ACS Nano (2014)}\\
         Tiling Phosphorene}

\abbreviations{DFT,PBE} %
\keywords{phosphorene, DFT, structure, band structure, stability}

\begin{document}

\begin{abstract}
We present
a scheme to categorize the structure of different layered
phosphorene allotropes by mapping their non-planar atomic
structure onto a two-color 2D triangular tiling pattern. In the
buckled structure of a phosphorene monolayer, we assign atoms in
``top'' positions to dark tiles and atoms in ``bottom'' positions
to light tiles. Optimum $sp^3$ bonding is maintained throughout
the structure when each triangular tile is surrounded by the same
number $N$ of like-colored tiles, with $0{\le}N{\le}2$. Our {\em
ab initio} density functional calculations indicate that both the
relative stability and electronic properties depend primarily on
the structural index $N$. The proposed mapping approach may also
be applied to phosphorene structures with non-hexagonal rings and
2D quasicrystals with no translational symmetry, which we predict
to be nearly as stable as the hexagonal network.
\end{abstract}

\vspace{10mm}

{\bf Keywords:} black phosphorus, phosphorene, DFT, {\em ab
initio}, structure, band structure, stability

\vspace{10mm}

Phosphorene, a monolayer of black phosphorus, is emerging as a
viable contender in the field of two-dimensional (2D) electronic
materials.\cite{Narita1983,Maruyama1981,DT229} In comparison to
the widely discussed semi-metallic graphene, phosphorene displays
a significant band gap while still maintaining a high carrier
mobility.\cite{Li2014,DT229,Koenig14,Xia14} The flexible structure
of semiconducting phosphorene\cite{Yang14,CastroNeto14} is
advantageous in applications including gas sensing\cite{Kou14},
thermoelectrics\cite{Fei14}, and Li-ion batteries.\cite{Fei14}
Unlike flat $sp^2$-bonded graphene monolayers, the structure of
$sp^3$-bonded phosphorene is buckled. There is a large number of
$sp^3$-bonded layered phosphorene structures, including blue-P,
$\gamma$-P, and $\delta$-P,\cite{DT230,DT232} which are nearly as
stable as the related black phosphorene structure but exhibit very
different electronic properties.
We believe that the above list of stable phosphorene structures is
still incomplete, giving rise to an unprecedented richness in
terms of polymorphs and their electronic structure.


\begin{figure*}[t]
\includegraphics[width=1.6\columnwidth]{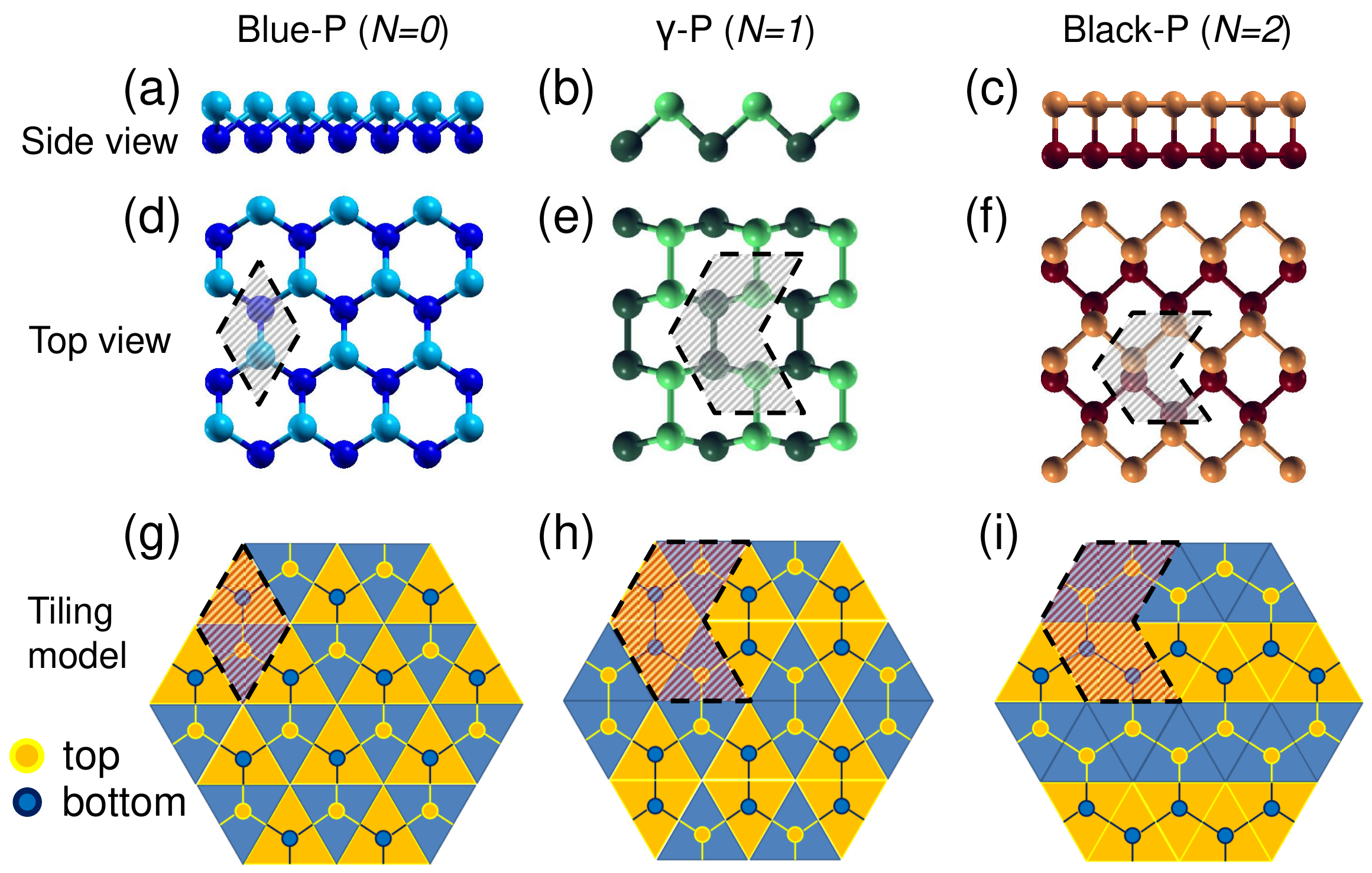}
\caption{(Color online) Atomic structure of different phosphorene
allotropes in (a-c) side view, (d-f) top view, and (g-i) in a
tiling model representation. Blue-P (a,d,g), $\gamma$-P (b,e,h),
and black-P (c,f,i) can be distinguished by the structural index
$N$. Primitive unit cells are emphasized by shading and delimited
by black dashed lines in (d-i). Atoms at the top and bottom of the
layer, as well as the corresponding tiles, are distinguished by
color.
\label{fig1}}
\end{figure*}

Here we introduce a scheme to categorize the structure of
different layered phosphorene allotropes by mapping the non-planar
3D structure of three-fold coordinated P atoms onto a two-color 2D
triangular tiling pattern. In the buckled structure of a
phosphorene monolayer, we assign atoms in ``top'' positions to
dark tiles and atoms in ``bottom'' positions to light tiles.
Optimum $sp^3$ bonding is maintained throughout the structure when
each triangular tile is surrounded by the same number $N$ of
like-colored tiles, with $0{\le}N{\le}2$. Our {\em ab initio}
density functional calculations indicate that both the relative
stability and electronic properties depend primarily on the
structural index $N$. Common characteristics of allotropes with
identical $N$ suggest the usefulness of the structural index for
categorization. The proposed mapping approach may also be applied
to phosphorene structures with non-hexagonal rings, counterparts
of planar haeckelite,\cite{DT228,Terrones2000} to point and line
defects,\cite{Yakobson14} and to 2D quasicrystals with no
translational symmetry, which we predict to be nearly as stable as
the hexagonal network.

\section{Results and discussion}

The non-planar atomic structure of selected $sp^3$-bonded
phosphorene allotropes is depicted in side and top view in
Fig.~\ref{fig1}(a-f). We find it convenient to map the 3D
structure of a phosphorene monolayer with threefold coordinated
atoms onto a 2D tiling pattern by assigning a triangular tile to
each atom, as shown in Fig.~\ref{fig1}(g-i). There is a one-to-one
correspondence between structures and tiling patterns, so that
different structures can be distinguished by different tiling
patterns. Dark-colored tiles are associated with atoms at the top
and light-colored tiles with atoms at the bottom of the layer.
Since each atom has 3 neighbors, each triangular tile is
surrounded by 3 neighboring tiles, $N$ of which have the same
color. It is obvious that $0{\le}N{\le}2$ provides the atom
associated with the central tile with a tetrahedral neighbor
coordination associated with the favorable $sp^3$ bonding. In our
tiling model, $N=3$ would represent the planar structure of an
energetically unfavorable $sp^2$-bonded lattice that, according to
our findings, would spontaneously convert to a non-planar
$sp^3$-bonded allotrope.

As we will show in the following, different allotropes with $N=0$,
$N=1$ and $N=2$ share similar characteristics. Therefore, the
structural index $N$ is useful for primary categorization of the
allotropes. In each structure depicted in Fig.~\ref{fig1}, $N$
maintains an identical value throughout the lattice, keeping the
favorable $sp^3$ bonding at all sites. We believe that this is the
underlying reason for our finding that these structures are nearly
equally stable.\cite{DT230,DT232}


\begin{figure}[t]
\includegraphics[width=1.0\columnwidth]{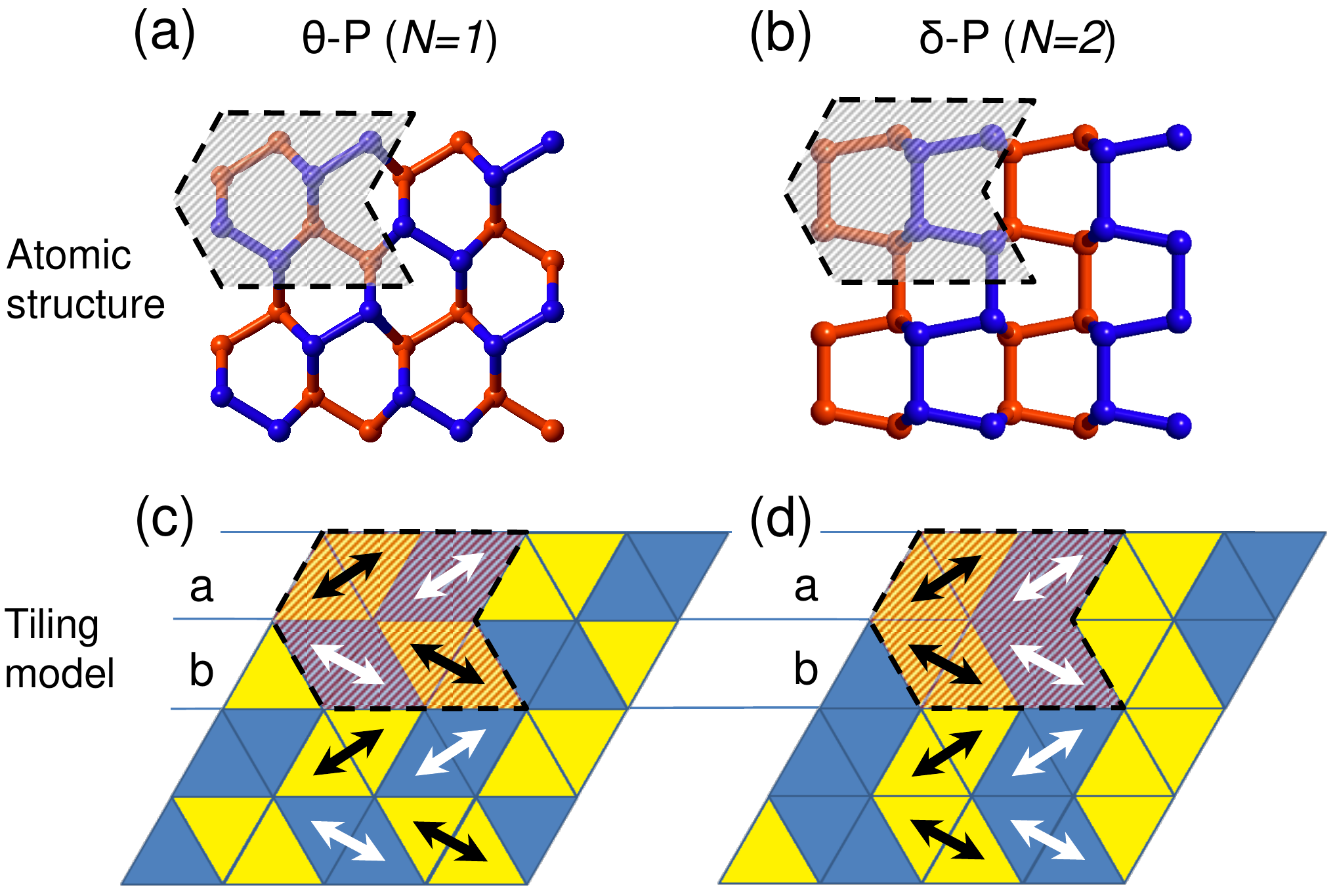}
\caption{(Color online) Atomic structure of $N=1$ and $N=2$
phosphorene allotropes in top view (a-b) and the corresponding
tiling model representation (c-d). The $N=1$ $\theta$-P allotrope
in (a,c) and the $N=2$ allotrope in (b,d) are structurally
different from the allotropes with the same $N$ in
Fig.~\protect\ref{fig1}. Primitive unit cells are emphasized by
shading and delimited by black dashed lines. Atoms at the top and
bottom of the layer, as well as the corresponding tiles, are
distinguished by color. The two different orientations of bonds
between like atoms, indicated by the double arrows as guides to
the eye, are denoted by ``$a$'' and ``$b$''. \label{fig2}}
\end{figure}

In the first category characterized by $N=0$, all neighbors of a
given atom have the same, but different height within the layer,
as seen in Figs.~\ref{fig1}(a) and \ref{fig1}(d). This translates
into a tiling pattern, where all adjacent tiles have a different
color, as seen in Fig.~\ref{fig1}(g). There is only one structural
realization within the $N=0$ category, namely the blue-P
allotrope.


In the second category characterized by $N=1$, each atom has one
like neighbor at the same height and two unlike neighbors at a
different height within the layer, as seen in Figs.~\ref{fig1}(b)
and \ref{fig1}(e). Besides the $\gamma$-P structure in
Fig.~\ref{fig1}(b) and \ref{fig1}(e), there is a $\theta$-P
allotrope, depicted Fig.~\ref{fig2}(a), with the same structural
index $N=1$. The tiling patterns of $\gamma$-P and $\theta$-P,
shown in Figs.~\ref{fig1}(h) and \ref{fig2}(c), are characterized
by a diamond harlequin pattern. Each diamond, formed of two
adjacent like-colored triangles, is surrounded by unlike-colored
diamonds. As a guide to the eye, we indicate the orientation of
the diamonds, same as the direction of the atomic bonds, by the
double arrows in Fig.~\ref{fig2}(c). The shape of the primitive
unit cells shown in Figs.~\ref{fig1} and \ref{fig2} is chosen to
see more easily the correspondence between the atomic structure
and the tiling pattern. The primitive unit cell of $\gamma$-P
contains 4 atoms according to Fig.~\ref{fig1}(h) and that of
$\theta$-P contains 8 atoms, as seen in Fig.~\ref{fig2}(c). As
indicated in Fig.~\ref{fig2}(c), the orientation of diamonds in a
row may be distinguished by the letters ``$a$'' or ``$b$''.
Whereas the perfect $\gamma$-P structure in Fig.~\ref{fig1}(h)
could be characterized by the sequence ``$aaaa\ldots$'' and the
structure of $\theta$-P by the sequence ``$abab\ldots$'', an
infinite number of different sequences including ``$abaa\ldots$''
would result in an infinite number of $N=1$ phosphorene
structures.


The most stable and best-known phosphorene allotrope is black-P,
depicted in Fig.~\ref{fig1}(c) and \ref{fig1}(f). Each atom in
this structure has two like neighbors at the same height and one
unlike neighbor at a different height, yielding a structural index
$N=2$. The tiling model of this structure type, shown in
Fig.~\ref{fig1}(i), contains contiguous arrays of like-colored
diamonds. These arrays may be either straight, as in
Fig.~\ref{fig1}(i) for black-P, or not straight, as in
Fig.~\ref{fig2}(d) for the structurally different $\delta$-P
allotrope with the atomic structure shown in Fig.~\ref{fig2}(b).
Describing diamond orientation by letters ``$a$'' and ``$b$'' as
in the case of $N=1$, we may characterize black-P in
Fig.~\ref{fig1}(i) by the sequence ``$aaaa\ldots$'' and $\delta$-P
in Fig.~\ref{fig2}(d) by the sequence ``$abab\ldots$''. As in the
case of $N=1$, an infinite number of different sequences including
``$abaa\ldots$'' would result in an infinite number of $N=2$
phosphorene structures.

\begin{figure*}[t]
\includegraphics[width=1.8\columnwidth]{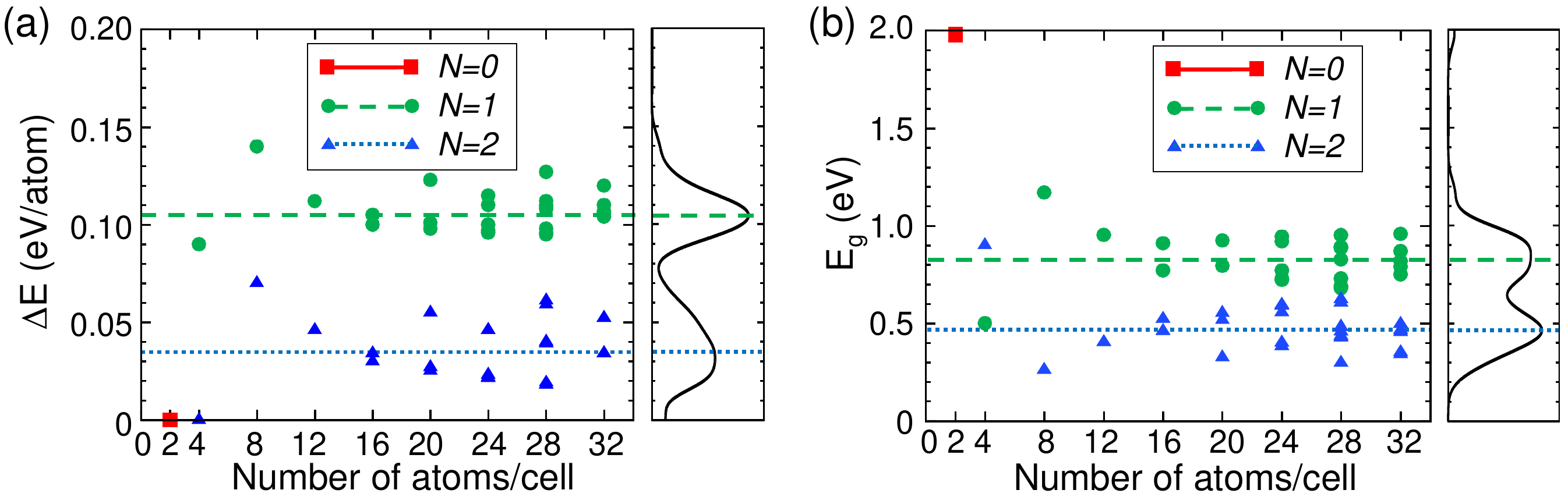}
\caption{(Color online) (a) Relative stability ${\Delta}E$ of
different phosphorene allotropes with respect to black phosphorus
and (b) their fundamental band gap $E_g$. The distribution of the
${\Delta}E$ and $E_g$ values, provided in the right panels of the
subfigures, indicates presence of three distinguishable groups
that may be linked to the different values of the structural index
$N$. The dashed and dotted lines are guides to the eye.
\label{fig3}}
\end{figure*}

The structural similarity and energetic near-degeneracy of $N=2$
and $N=1$ structures stems from the fact that a structural change
from $N=2$ to $N=1$ involves only a horizontal shift of every
other row, indicated by the horizontal lines in
Figs.~\ref{fig2}(c) and \ref{fig2}(d), by one tile. It is even
possible to generate structural domains with different values of
$N$. The energy cost of domain wall boundaries may be extremely
low\cite{DT232} if optimum $sp^3$ bonding is maintained at the
boundaries.

As mentioned above, there is only one allotrope with $N=0$, but
infinitely many structures with $N=1$ and $N=2$. Of these, we
identified and optimized all lattices with up to 28 atoms per
primitive unit cell and selected other structures with up to 32
atoms per unit cell. For each lattice, we identified the relative
stability ${\Delta}E$ with respect to the most stable black
phosphorene allotrope on a per-atom basis and plotted the values
in Fig.~\ref{fig3}(a).

The electronic band structure of systems with large unit cells is
very dense and hard to interpret in comparison to that of the
allotropes discussed in Figs.~\ref{fig1} and \ref{fig2}, which is
reproduced in the Supporting Information.\cite{Ptile14-SM} For
each of these structures, though, we identified the value $E_g$ of
the fundamental band gap and provide the results in
Fig.~\ref{fig3}(b).

We find that neither ${\Delta}E$ nor $E_g$ display a general
dependence on the size of the unit cell. We found all structures
to be relatively stable. The small values ${\Delta}E<0.15$~eV/atom
indicate a likely coexistence of different allotropes that would
form under nonequilibrium conditions. All band gap values, which
are typically underestimated in DFT-PBE
calculations,\cite{{SIESTA},{PBE}} occur in the range between
$0.3$~eV and $2.0$~eV, similar to the allotropes discussed in
Figs.~\ref{fig1} and \ref{fig2}. Rather surprisingly, the
distribution of ${\Delta}E$ and $E_g$ values, shown in the right
panels of the sub-figures Fig.~\ref{fig3}(a) and \ref{fig3}(b),
exhibits three peaks that can be associated with the structural
index $N$, with a rather narrow variance caused by the differences
between the allotrope structures. We find the energetically
near-degenerate blue phosphorene ($N=0$ with 2 atoms per unit
cell) and black phosphorene ($N=2$ with 4 atoms per unit cell)
structures to be the most stable, followed by other $N=2$
structures with more than 4 atoms per unit cell. We found $N=1$
structures to be the least stable of all. Similarly, the $N=0$
blue phosphorene allotrope has the largest band gap, $N=2$
allotropes have the smallest band gap, and $N=1$ allotropes lie in
between.




\begin{figure*}[t]
\includegraphics[width=1.8\columnwidth]{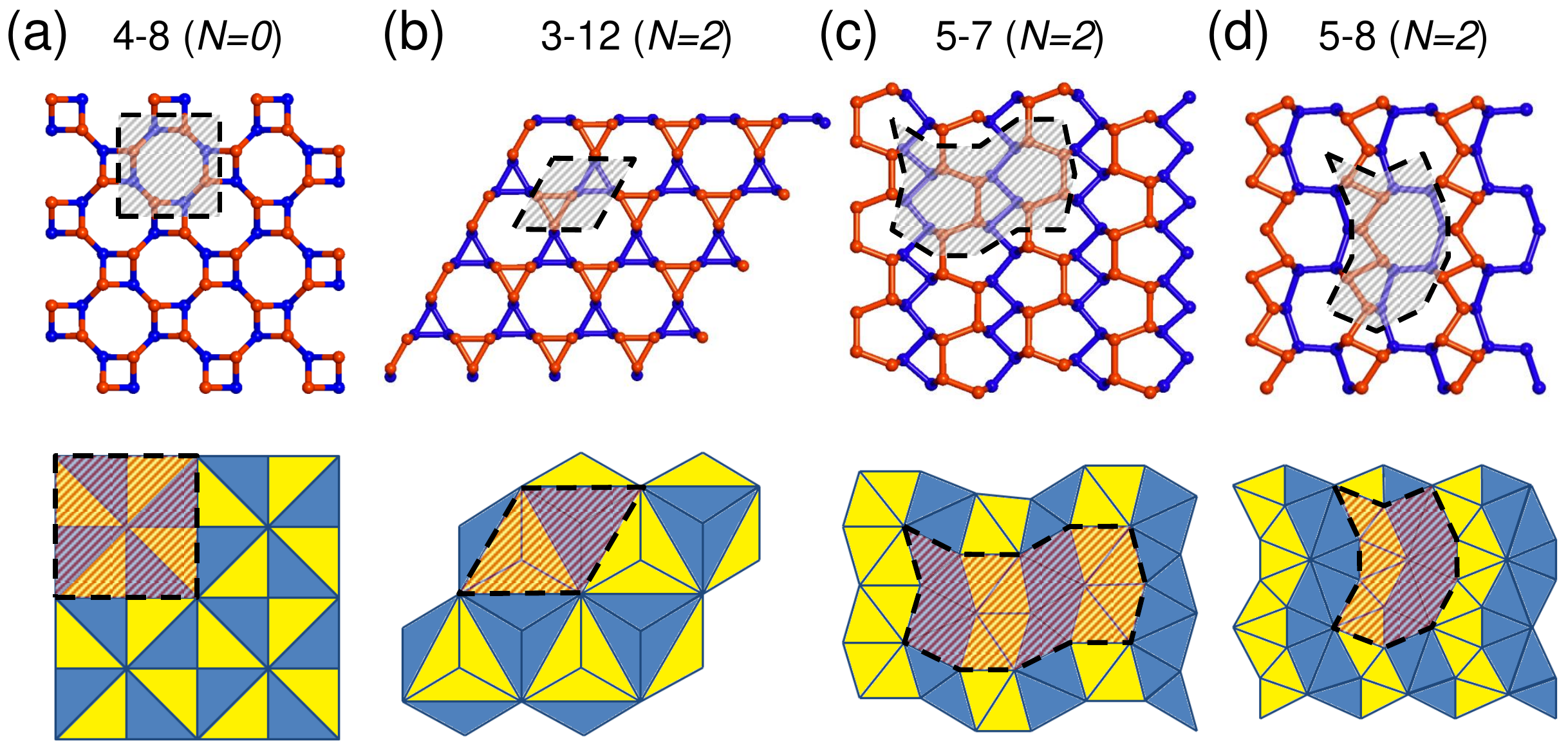}
\caption{(Color online) Equilibrium structures (top) and
corresponding tiling patterns (bottom) of 2D phosphorene with (a)
4-8, (b) 3-12, (c) 5-7, and (d) 5-8 membered rings. The color of
the atoms and corresponding tiles distinguishes positions at the
top and the bottom of the layer. \label{fig4}}
\end{figure*}

The higher stability of $N=2$ phosphorene structures in comparison
to $N=1$ allotropes indicates an energetic preference for
phosphorus atoms forming zigzag chains at the same height rather
than forming isolated dimers. Among the $N=2$ structures,
$\delta$-P is the least stable, with
${\Delta}E{\approx}0.07$~eV/atom. All the other $N=2$ structures
fall in-between $\delta$-P and black phosphorus in terms of
stability. This finding is easy to understand, since all these
structures are combination of black phosphorus and $\delta$-P.

For both $N=2$ and $N=1$ allotropes, we find structures with the
same orientation of diamonds in the tiling pattern to be more
stable. The $\gamma$-P structure, with all diamonds aligned in the
same direction in the tiling pattern, is the most stable $N=1$
phosphorene allotrope, but still less stable by $0.09$~eV/atom
than the $N=2$ black phosphorene. At the other extreme of the
relative stability range, $\theta$-P with disordered diamond
orientations in the tiling pattern is the least stable $N=1$
allotrope, being $0.14$~eV/atom less stable than black
phosphorene. In analogy to what we concluded for $N=2$ structures,
all $N=1$ phosphorene allotropes can be viewed as a combination of
$\gamma$-P and $\theta$-P, with their stability in-between the
above limiting values.

As mentioned above, also the distribution of $E_g$ values, shown
in the right panel of Fig.~\ref{fig3}(b), indicates three distinct
groups that can be associated with the structural index $N$. The
largest band gap value of $2.0$~eV
in the only $N=0$ structure, blue phosphorene, is well separated
from the band gap distribution of $N=1$ and $N=2$ structures that
form a double-hump shape. We note that the two peaks in the band
gap distribution of $N=1$ and $N=2$ allotropes are not as well
separated as the two peaks in the stability distribution in
Fig.~\ref{fig3}(a), so the trends in the band gap value are not as
clear as trends in the relative stability. In systems with large
unit cells, band gaps of $N=1$ structures are grouped around
$0.8$~eV, whereas band gaps of $N=2$ structures are grouped around
$0.5$~eV. The largest spread in $E_g$ values is in systems with
very small unit cells. Among $N=1$ allotropes, we find the
smallest value $E_g{\approx}0.5$~eV in the structure with 4
atoms/unit cell ($\gamma$-P) and the largest value
$E_g{\approx}1.2$~eV
in the structure with 8 atoms/unit cell ($\theta$-P). Band gap
values of other $N=1$ structures range between these two values.
$N=2$ structures have generally the lowest band gap values of the
three groups. Among $N=2$ systems, we find the largest value
$E_g{\approx}0.9$~eV in the structure with 4 atoms/unit cell
(black phosphorene) and $E_g{\approx}0.3$~eV
in a system with 8 atoms/unit cell, the smallest gap value among
several metastable structures of $\delta$-P. Band gap values of
other $N=2$ structures range between these two values. As
discussed earlier,\cite{DT230,DT232} our PBE-based band gap values
are generally underestimated. More precise quasiparticle
calculations beyond DFT, including the GW formalism, indicate that
the band gap values should be about $1$~eV larger than the PBE
values presented here.\cite{DT230,YangGW}

As the unit cell size of $N=1$ and $N=2$ structures grows
infinitely large, we gradually approach amorphous phosphorene.
Assuming that our findings in Fig.~\ref{fig3} are universal and
not limited to the finite sizes addressed by our study, we
conclude that the stability and the fundamental band gap of such
amorphous structures should also be found in the range suggested
by their structural index $N$.


The one-to-one mapping between 3D structures of periodic systems
and 2D tiling patterns is not limited to a honeycomb lattice with
6-membered rings, but can equally well be applied to lattices with
$3-$, $4-$, $5-$, $7-$, $8-$ and $12-$membered rings found in
planar haeckelites.\cite{DT228,Terrones2000} The corresponding
geometries and tiling patterns are shown in Fig.~\ref{fig4}. Among
these structures, 4-8 phosphorene has the highest symmetry, a
relatively small unit cell with the shape of a square and a tiling
pattern composed of right triangles. Besides the $N=0$ structure
depicted in Fig.~\ref{fig4}(a), we can identify allotropes with
4-8 rings with structural indices $N=1$ and $N=2$. Other
allotropes with 3-12, 5-7 and 5-8 rings, shown in
Figs.~\ref{fig4}(b-d), may not exist in all the variants of the
structural index $N$ due to their lower symmetry. For example, the
allotrope with 5-7 rings does not have a structure with $N=0$.

We find structures with non-hexagonal rings to be generally less
stable than the most stable black phosphorene, but the energy
differences ${\Delta}E<0.2$~eV/atom are very small.
Consequently, we expect that such structures should coexist with
black phosphorene as either pure phases, or as local defects at
domain wall boundaries, or as finite-size domains in the host
layer. We find all phosphorene allotropes with non-hexagonal rings
to be semiconducting, with the band gap determined primarily by
the structural index $N$.

\begin{figure}[t]
\includegraphics[width=1.0\columnwidth]{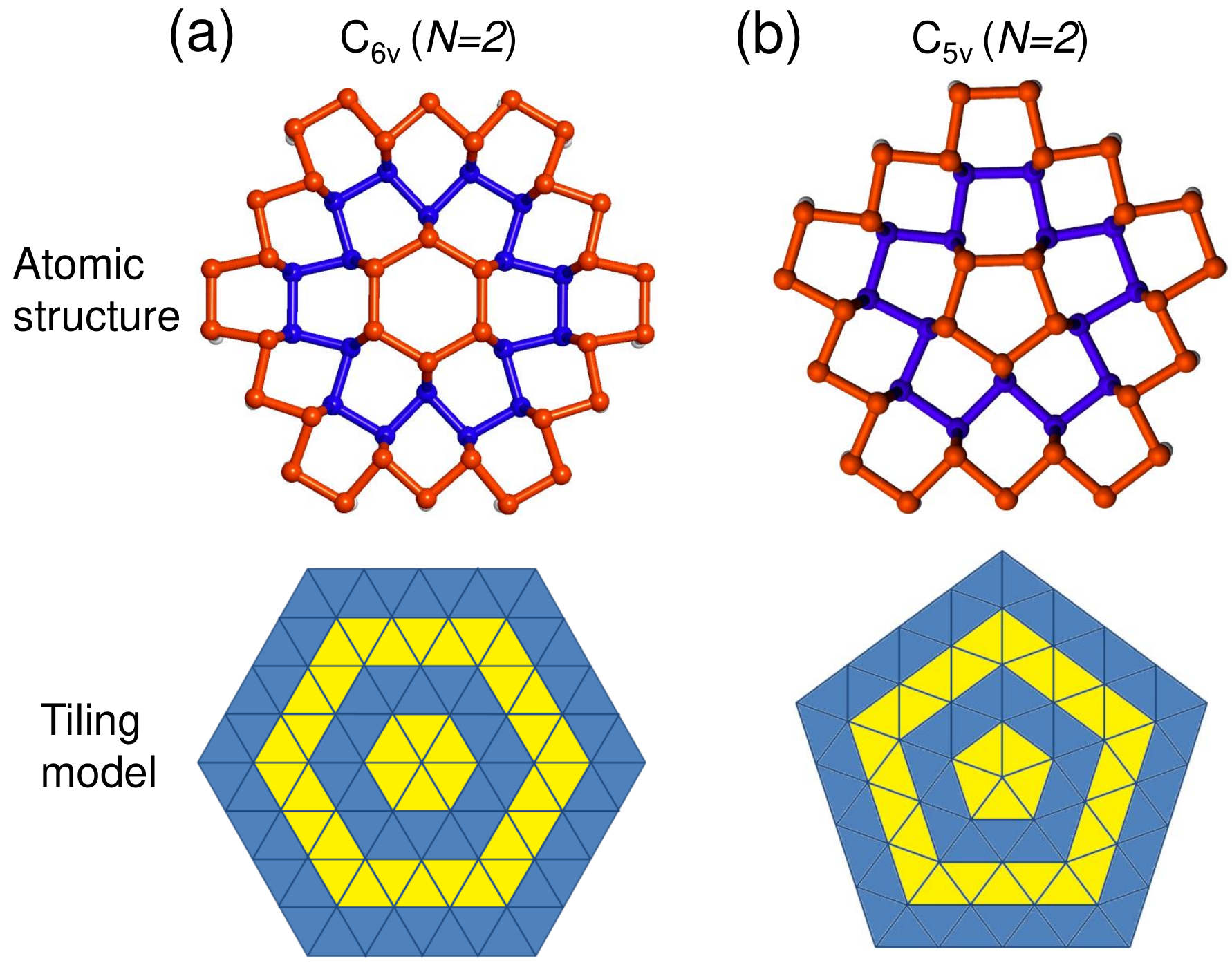}
\caption{(Color online) Equilibrium structures (top panels) and
corresponding tiling patterns (bottom panels) of phosphorene
structures with (a) a $C_{6v}$ and (b) a $C_{5v}$ symmetry. Edges
of the finite-size flakes are terminated by hydrogen atoms,
colored in white. The color of the phosphorus atoms and the
corresponding tiles distinguishes positions at the top or the
bottom of the layer. \label{fig5}}
\end{figure}

Phosphorene may also form aperiodic structures with no
translational symmetry. Examples of such systems with only
rotational symmetry are shown in Fig.~\ref{fig5}.
Fig.~\ref{fig5}(a) depicts a phosphorene structure of type $N=2$
with a $C_{6v}$ point group symmetry and the corresponding tiling
pattern. In this structure, arrays of neighboring atoms form an
alternating circular pattern about the center that can cover an
infinite plane. The analogous $N=2$ structure with $C_{5v}$
symmetry is depicted in Fig.~\ref{fig5}(b), and analogous
structures with $C_{nv}$ symmetry could be imagined as well. To
judge the stability of these aperiodic structures, we optimized
finite-size flakes that were terminated by hydrogen atoms at the
exposed edge. We found these structures to be semiconducting and
as stable as the periodic structures discussed in
Fig.~\ref{fig3}(a), with ${\Delta}E=0.07$~eV/atom for the $C_{6v}$
and ${\Delta}E=0.04$~eV/atom for the $C_{5v}$ structure falling
into range expected for $N=2$.
%
%
%
%
%
%


These findings indicate that our classification scheme and tiling
model is useful to characterize monolayers of three-fold
coordinated, $sp^3$-hybridized phosphorus atoms arranged in
periodic or aperiodic patterns. Due to structural similarities
between layered structures of group-V elements, we believe that
our findings regarding relative stability, electronic structure
and fundamental band gap will likely also apply to other systems
including monolayers of arsenic, antimony and bismuth.

Since the cohesive energy differences are rather small, we must
consider the possibility that the stability ranking of the
different allotropes at $T=0$ and related properties\cite{Han2014}
may depend on the DFT functional. We have compared PBE results for
the relative stability of the different allotropes with LDA
results and found the maximum difference in the relative
stabilities of the different allotropes to be $0.02$~eV/atom,
which does not change the energy ranking of the allotropes.

Since phosphorene structures will likely be synthesized at nonzero
temperatures, the relative abundance of different allotropes will
depend on their free energy at that temperature. Consequently, our
total energy results for stability differences at $T=0$ need to be
corrected by also addressing differences in entropy at $T>0$. Even
though the decrease in free energy with increasing temperature
should be similar in the different allotropes due to their similar
vibration spectra,\cite{{DT230},{DT232},{Ptile14-SM}} minute
differences in vibrational entropy may become important in view of
the small differences between stabilities of the allotropes at
$T=0$, and could eventually change the free energy ranking at high
temperatures.


In conclusion, we have introduced a scheme to categorize the
structure of different layered phosphorene allotropes by mapping
the non-planar 3D structure of three-fold coordinated P atoms onto
a two-color 2D triangular tiling pattern. In the buckled structure
of a phosphorene monolayer, we assign atoms in ``top'' positions
to dark tiles and atoms in ``bottom'' positions to light tiles. We
found that optimum $sp^3$ bonding is maintained throughout the
structure when each triangular tile is surrounded by the same
number $N$ of like-colored tiles, with $0{\le}N{\le}2$. Our {\em
ab initio} density functional calculations indicate that both the
relative stability and electronic properties depend primarily on
the structural index $N$. Common characteristics of allotropes
with identical $N$ suggest the usefulness of the structural index
for categorization. The proposed mapping approach may also be
applied to phosphorene structures with non-hexagonal rings and to
2D quasicrystals with no translational symmetry, which we predict
to be nearly as stable as the hexagonal network.

\section{Methods}

Our computational approach to gain insight into the equilibrium
structure, stability and electronic properties of various
phosphorene structures is based on {\em ab initio} density
functional theory (DFT) as implemented in the
\textsc{SIESTA}\cite{SIESTA}. We used periodic boundary conditions
throughout the study. We used the Perdew-Burke-Ernzerhof
(PBE)\cite{PBE} exchange-correlation functional, norm-conserving
Troullier-Martins pseudopotentials\cite{Troullier91}, and a
double-$\zeta$ basis including polarization orbitals. Selected PBE
results were compared to results based on the Local Density
Approximation (LDA)~\cite{{Ceperley1980},{Perdew81}}. The
reciprocal space was sampled by a fine grid\cite{Monkhorst-Pack76}
of $8{\times}8{\times}1$~$k$-points in the Brillouin zone of the
primitive unit cell. We used a mesh cutoff energy of $180$~Ry to
determine the self-consistent charge density, which provided us
with a precision in total energy of $\leq$ 2~meV/atom. All
geometries have been optimized by \textsc{SIESTA} using the
conjugate gradient method\cite{CGmethod}, until none of the
residual Hellmann-Feynman forces exceeded $10^{-2}$~eV/{\AA}.

\begin{acknowledgement}
We thank Luke Shulenburger for useful discussions. This study was
supported by the National Science Foundation Cooperative Agreement
\#EEC-0832785, titled ``NSEC: Center for High-rate
Nanomanufacturing''. Computational resources have been provided by
the Michigan State University High Performance Computing Center.
shown in this document.
\end{acknowledgement}


\begin{suppinfo}
Electronic band structure of the phosphorene allotropes discussed
in Figs.~\protect\ref{fig1} and \protect\ref{fig2} and phonon band
structure of $\theta$-P.
\end{suppinfo}


\providecommand*\mcitethebibliography{\thebibliography} \csname
@ifundefined\endcsname{endmcitethebibliography}
  {\let\endmcitethebibliography\endthebibliography}{}

\end{document}